\documentclass[
aps,
prb,
twocolumn,
superscriptaddress,%
showpacs,%
preprintnumbers,%
byrevtex,%
floatfix%
]{revtex4}

\usepackage{dcolumn}% Align table columns on decimal point
\usepackage{bm}% bold math
\usepackage{ifthen}
\usepackage[usenames]{color}
\usepackage{amssymb}
\usepackage{amsfonts}
\usepackage{amsmath}
\usepackage[dvips]{graphicx}
\begin{document}

\date{\today}
\title{Mechanisms for p-type behavior of ZnO, Zn$_{1-x}$Mg$_x$O and related oxide semiconductors}
\author{Daniel F. Urban}
\email{daniel.urban@iwm.fraunhofer.de}
\affiliation{Fraunhofer Institute for Mechanics of Materials IWM, W\"ohlerstr. 
11, 79108 Freiburg, Germany}

\author{Wolfgang K\"orner}
\email{wolfgang.koerner@iwm.fraunhofer.de}
\affiliation{Fraunhofer Institute for Mechanics of Materials IWM, W\"ohlerstr.
11, 79108 Freiburg, Germany}
\author{Christian Els\"asser}
\affiliation{Fraunhofer Institute for Mechanics of Materials IWM, W\"ohlerstr. 
11, 79108 Freiburg, Germany}
\affiliation{University of Freiburg, Freiburg Materials Research Center (FMF), 
Stefan-Meier-Str. 21, 79104 Freiburg, Germany}

\begin{abstract}
Possibilities of turning intrinsically n-type oxide semiconductors like ZnO and 
Zn$_{1-x}$Mg$_x$O into p-type materials are investigated.
Motivated by recent experiments on Zn$_{1-x}$Mg$_x$O doped with nitrogen we 
analyze the electronic defect levels of point defects N$_{{\rm O}}$, v$_{{\rm 
Zn}}$, and N$_{{\rm O}}$-v$_{{\rm Zn}}$ pairs in ZnO and Zn$_{1-x}$Mg$_x$O by 
means of self-interaction-corrected density functional theory calculations.
We show how the interplay of defects can lead to shallow acceptor defect 
levels, although the levels of individual point defects N$_{{\rm O}}$ are too 
deep in the band gap for being responsible for p-type conduction. We relate our 
results to p-type conduction paths at grain boundaries seen in polycrystalline 
ZnO and develop an understanding of a p-type mechanism which
is common to ZnO, Zn$_{1-x}$Mg$_x$O, and related materials.
\end{abstract}

%band structure of, 71.20.Mq, 71.20.Nr
%impurity and defect levels, 71.55.Jv
%electron density of states, 71.23.-k

\maketitle
\section{Introduction}

Zinc oxide is a prominent example of an intrinsically n-type semiconductor.
Its abundant availability, its low-cost production, and its wide electronic
band gap of approximately 3.4 eV make ZnO attractive for transparent and 
conducting oxide layers in consumer-electronics devices.
However, a break-through of ZnO-based transparent electronics has not been
achieved so far due to the lack of sufficiently conductive and long-term stable
\hbox{p-type} ZnO layers. The problem of p-type doping of ZnO has by now 
been a scientific challenge for more than two decades.
Besides the doping with single elements like N, P, As, or Sb, substituting O 
and Li, Na, K, Cu, Ag or Au, substituting Zn\cite{le01,le04,le06,pa02,li04,ya08} 
co-doping was discussed as well.\cite{wa03,zha06,ya07}
Apparently, the most promising candidate so far is still nitrogen. However, it
was shown theoretically\cite{ly09,la10,ko10} that the substitutional defect of 
N on O sites (N$_{{\rm O}}$)  cannot be responsible for the p-type conduction 
which has been observed in some experiments,\cite{re13} since it creates a 
defect level which is located too high above the valence band (VB) edge with 
respect to thermal excitation energies.

Recent theoretical and experimental studies consider more complicated defect
complexes which involve zinc vacancies (v$_{{\rm Zn}}$) or co-doping, like the 
N$_{{\rm O}}$-v$_{{\rm Zn}}$ or N$_{{\rm O}}$-v$_{{\rm Zn}}$-H defect 
complexes, as potential sources of p-type conduction.\cite{liu12,am15}
 
For polycrystalline ZnO it has been proposed that conduction channels of p-type 
can be found along grain boundaries.\cite{wa09,ko10} Here it is an interplay 
of grain boundaries, their depletion regions, and point defects that
results in the desired low lying acceptor levels.

It has been reported by Hierro et al.\cite{hi09} that the use of ternary ZnO-based alloys like
Zn$_{1-x}$Mg$_x$O (denoted by ZMO in the following) facilitates the p-type 
conductivity due to lower net electron concentrations than in pure 
ZnO. 
These authors relate the reduced net electron concentration to two deep lying Mg-related defects which act as electron traps.
ZMO retains the wurtzite crystal structure of ZnO for Mg 
substituting Zn for concentrations up to $x \approx 40\%$, and also the
lattice parameters change only slightly. Hence combining ZMO with n-type ZnO 
layers  leads to negligible lattice misfits.\cite{hi09}

Up to now it is not yet clear which atomic or extended defects provide the key 
to promote p-type conductivity in ZnO and related materials like ZMO.
An understanding of this defect physics would allow a targeted development of
experimental routes to synthesize  materials having these particular acceptor 
states.

In this paper we present a density functional theory (DFT) study determining
the electronic defect levels of the defects N$_{{\rm O}}$, v$_{{\rm Zn}}$, and 
N$_{{\rm O}}$-v$_{{\rm Zn}}$ in ZnO and ZMO. We discuss our results in the
context of the accumulated experimental and theoretical knowledge, and thereby 
we shed light on the question of why some experimental routes recently taken 
have lead to p-conduction but others not.
Are the defect states that originate from N$_{{\rm O}}$ indeed too
deep\cite{ly09,ko10} in the band gap?
What changes when N$_{{\rm O}}$ forms a defect complex with a Zn
vacancy, and how do Mg atoms nearby affect its defect level position?
For the determination of the electronic defect levels we use
a self-interaction corrected (SIC) local-density approximation (LDA)
approach\cite{ko10} that is capable to give far better results than both
commonly used approximations, the LDA and the generalized gradient
approximation (GGA). The latter two have limited predictive power for wide
band gap semiconductors due to their inherent significant underestimation of the
band gap due to their self-interaction error.
SIC--LDA overcomes this limitation and yields electronic band structures
with band gaps that are in good agreement with experiment\cite{ko10,ko11}
 and in particular for ZnO
a correct separation of the Zn-3d and O-2p bands.\cite{ko14}
By means of this approach, the defect levels in various amorphous
oxide semiconductors based on In, Sn, and Zn have been predicted
reliably.\cite{sa14,sa15,ko13,ko15}

The results presented in this paper may solve the controversial problem that 
individual N$_{{\rm O}}$ defects are indeed creating levels lying too deep 
inside the band gap even though nitrogen doping has been observed to lead 
to p-type ZnO and ZMO, probably due to an interplay with other defects. 
Moreover, we relate our present results to the p-type conduction effects which 
have been observed for grain boundaries in ZnO, and we relate all this to one common origin. 
The emerging theoretical picture  will presumably be useful for the
experimental optimization of p-type conduction in ZnO and ZMO.

\begin{figure}[]
      \begin{center}
      \includegraphics[width=4cm]{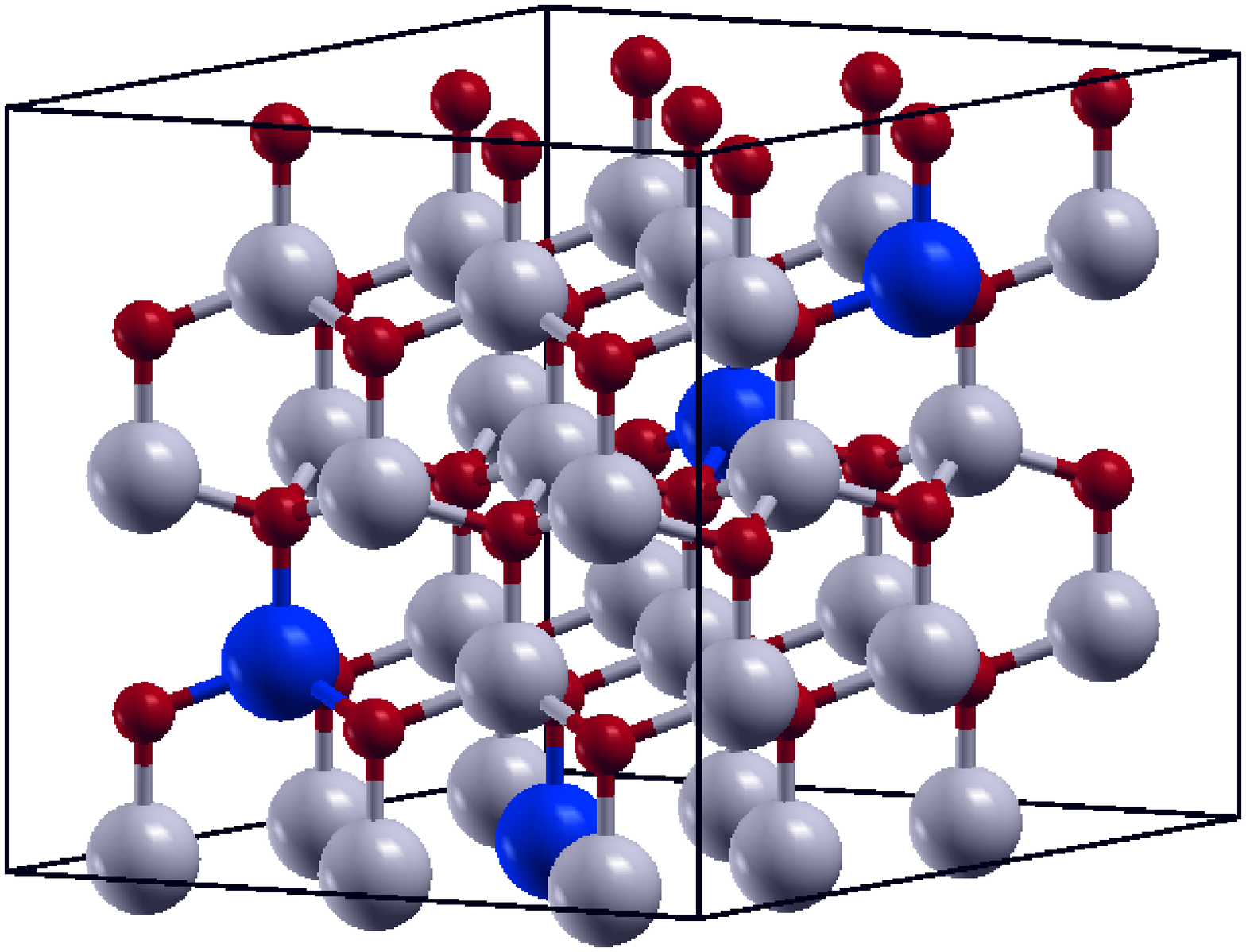}
      \includegraphics[width=4cm]{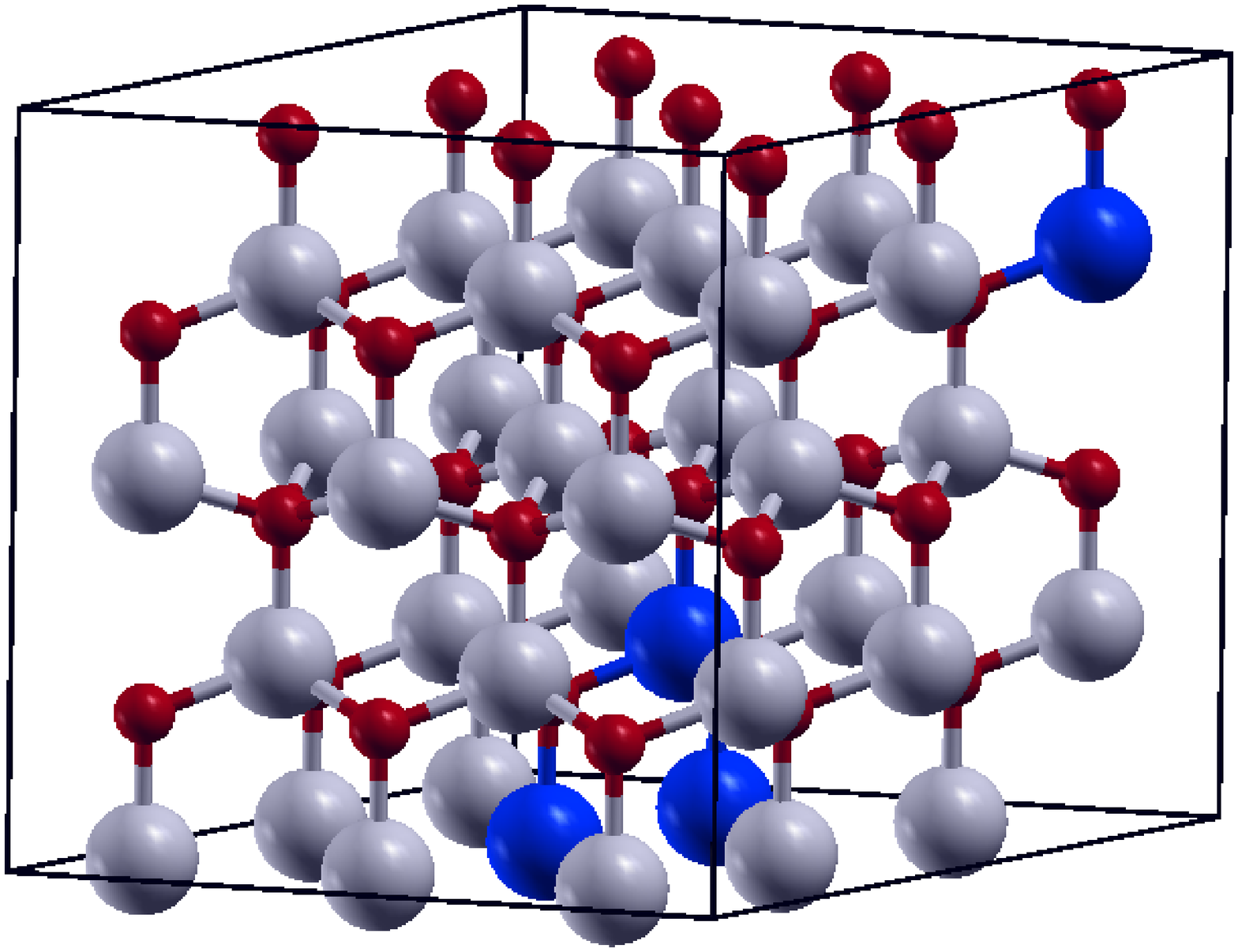}
\caption{(Color online)
Two examples of supercell models  Zn$_{32}$Mg$_4$O$_{36}$ with randomly
distributed Mg atoms (left) and with clustered Mg atoms (right). Large gray and
blue spheres represent Zn and Mg atoms, respectively.
Small red spheres represent oxygen atoms.
\label{fig:znmgo:cells}}
\end{center}
\end{figure}

\section{Computational approach}

\subsection{Supercell models}

For the study of defect complexes we have chosen a supercell model of the 
wurtzite crystal structure of ZnO containing 72 atoms and consisting of  
$3\times 3\times 2$ primitive unit cells. Since Zn$_{1-x}$Mg$_x$O retains the 
wurtzite structure for Mg concentrations $x\leq40$\%\cite{hi09} the 
corresponding supercells were generated by randomly substituting Zn atoms by Mg 
atoms. For the study of the bulk crystal properties like lattice parameters and 
band gap as function of the Mg concentration we compared several realizations 
of random Mg distributions within the 72-atoms supercell, and we find a 
negligible dependency of these properties on the Mg distribution. For 
comparison, we have also examined results from a 36-atom supercell in the study 
of bulk properties. For the subsequent study of defects we have chosen the 
specific Mg concentration of $x=1/9\sim$11\%  (i.e. replacing 4 of 36 Zn atoms 
by Mg atoms in the 72-atom supercell) which is representative for typical 
experimental concentrations.\cite{ku14} Furthermore, the defect analysis 
concentrates on specific realizations of the Mg distribution which differ in 
the number of Mg atoms that cluster in the direct neighborhood of the $N_{{\rm 
O}}$-point defect under study. Two such realizations are displayed in Fig.\ 
\ref{fig:znmgo:cells}.

\subsection{Structure optimization}

The structure optimization of all the constructed supercell models was carried 
out using the projector augmented-wave (PAW) method\cite{bl94} as implemented 
in the VASP code.\cite{kr96,kr99} The LDA was used for the exchange-correlation 
functional, and PAW potentials describing the Zn-(3d,4s,4p), Mg-3s, and 
O-(2s,2p) as valence electrons. All VASP calculations were carried out with a 
plane-wave cutoff energy of 400 eV, a $3\times 3\times 3$ Monkhorst-Pack k-mesh, 
and a Gaussian broadening of 0.1 eV.

For all bulk structures the volume optimization was carried out by fitting the
universal equation of state\cite{ro81} to a dataset of minimal total energies
calculated at different cell volumes. Optimization of the lattice parameter 
ratio c/a and internal relaxation of the atomic positions were achieved by 
minimizing the elastic stress and the forces acting on the atoms, respectively.
The supercells containing defects were internally relaxed at fixed volume and
c/a ratio.

\subsection{Electronic-structure calculations}

The electronic-structure calculations on the basis of the SIC-LDA were 
performed using the computational mixed-basis pseudopotential 
(MBPP) method\cite{el90,el92,lech04,mey95} with the same calculation setup
as reported in previous papers.\cite{ko10,ko11b,ko11c} We have taken the
LDA for exchange-correlation as parametrized by Perdew and Zunger.\cite{pe81} 
For Zn, Mg, and O optimally smooth norm--conserving pseudopotentials\cite{va85} 
were constructed, and a mixed basis of plane waves and non-overlapping 
localized orbitals were used. Due to the localized orbitals a plane--wave 
cutoff energy of 20 Ry (1Ry = 13.606 eV) is sufficient for obtaining well 
converged results. For the k-point sampling of the Brillouin-zone integrals a 
Monkhorst-Pack mesh of $3\times 3\times 3$ and a Gaussian broadening of 0.2 eV 
were used. The DOS of the supercells were evaluated with the same mesh and a 
Gaussian broadening of 0.1 eV.

The self-interaction of the LDA is corrected by an incorporation of the SIC in 
the pseudopotentials.\cite{vo96,ko10} The SIC procedure uses weight factors 
{\boldmath$w$}$=(w_s, w_p, w_d)$ accounting for the occupations of the
individual s, p and d valence orbitals. We corrected the Zn 3d 
semicore-orbitals by 100\%, i.e.\ {\boldmath$w$}$_{\rm Zn}$=(0, 0, 1).
The localized O-2s semi-core orbitals are also corrected  by 100\% while for 
the spatially more extended O-2p valence-band orbitals we have taken 80\% which 
implies the weight factors {\boldmath$w$}$_{\rm O}$=(1, 0.8, 0). We did not 
apply a SIC to the almost unoccupied Mg-3s, 3p and Zn-4s, 4p conduction-band 
orbitals of the almost fully ionic compounds Zn$_{1-x}$Mg$_x$O.
For the adjustment of the atomic SIC to the crystal field in the solids, 
a scaling factor $\alpha=0.8$ is applied.

\section{Results and discussion}

\subsection{Lattice parameters}

\begin{figure}[]
  \centerline{\includegraphics[width=8.5cm]{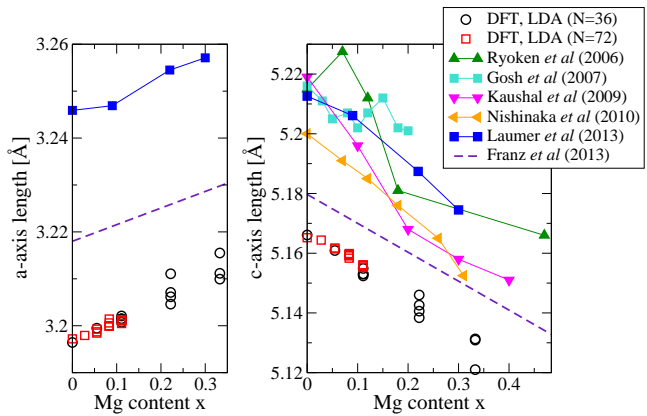}}
\caption{(Color online)
Lattice parameters obtained by DFT structure optimization
of ZMO supercells with N=36 (open circles) and N=72 (open squares) atoms
as function of the Mg content in comparison to experimental
data\cite{ry06,gh07,kau09,nish10,lau13} (filled symbols) and the result of a 
recent DFT study\cite{franz13} (dashed line), taken from the literature. 
Experimental data points are connected by straight solid lines to guide the 
eyes.
\label{fig:a_c}
}
\end{figure}

The functional dependencies of the lattice parameters $a$ and $c$ on the Mg
content are displayed in Fig.\  \ref{fig:a_c}. Our results underestimate the
experimental values by about one percent, which is a typical systematic
deviation for LDA results. The increase of $a$ and the decrease of $c$ with 
increasing Mg content are quantitatively correct and just slightly shifted with 
respect to previous experimental\cite{ry06,gh07,kau09,nish10,lau13}
and theoretical\cite{franz13} findings. This good agreement on the structural 
parameters is the prerequisite for the electronic structure calculations which 
are the focus of this work. Note that the deviations from  Vegard's rule 
in some of the values of the experimentally observed lattice constants is 
probably related to the quality of the thin films and maybe due to the
presence of compensated defects or residual strain as discussed, e.g., by Ryoken 
\emph{et al.}\cite{ry06}

\begin{figure}[]
  \centerline{\includegraphics[width=8.5cm]{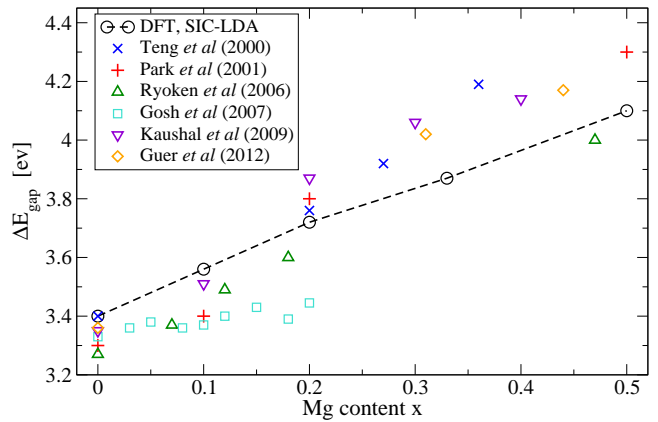}}
\caption{(Color online)
Band gap of Zn$_{1-x}$Mg$_x$O as function of the Mg content $x$. The data points
calculated with SIC-LDA are connected by a black dashed line to guide the eyes. 
Experimental data from several groups are shown for 
comparison.\cite{te00,pa01,ry06,gh07,gu12}
\label{fig:bandgap}
}
\end{figure}

\subsection{Band gap of ZMO}

Theoretical and experimental values of the band gap of
Zn$_{1-x}$Mg$_x$O as function of the Mg content $x$ are displayed and compared 
in Fig.\ \ref{fig:bandgap}. The results obtained by the SIC-LDA approach
do not only show the rising trend correctly, but also the calculated values 
agree quantitatively well to the experimental data. The dependency of the band 
gap on the specific distribution of Mg atoms within the supercell is small, 
e.g. for the sample structures with $x\sim22\%$ we find an energy scatter of 
only $\delta E\sim10$meV. 

The wide scatter of experimental results in Fig.\
\ref{fig:bandgap} is probably due to the varying quality of the thin films but 
also due to different experimental approaches of determining the band 
gaps.\cite{te00,pa01,ry06,gh07,gu12} Above the thermodynamical solubility limit 
of $x\sim5\%$ a non-equilibrium solid solution is formed. Depending on the 
route of synthesis, this results in compositional non-uniformity which then 
leads to the generation of non-equilibrium compensated defects\cite{gh07} or 
even structural inhomogeneity\cite{ry06} and residual elastic stress. 
Naturally, all of this is reflected in the measured physical properties of the 
films, as discussed in the respective articles.

For our investigation the calculation of the band gap as function of the Mg
content serves as a validation that the SIC-LDA describes
the electronic structure of the Zn-Mg-O system rather well.

\begin{figure}[]
\centerline{\includegraphics[width=8.5cm]{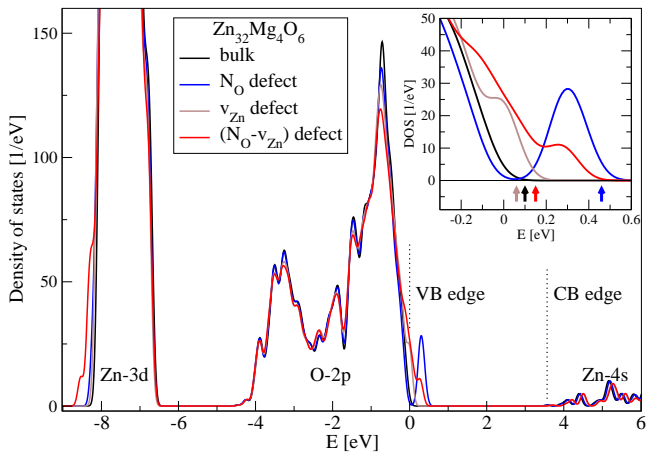}}
\caption{(Color online)
  Comparison of the total DOS of Zn$_{32}$Mg$_4$O$_{36}$ without (bulk) and with
  N$_{{\rm O}}$, v$_{{\rm Zn}}$,  and N$_{{\rm O}}$-v$_{{\rm Zn}}$ defects.
  The inset magnifies the region close to the VB maximum.  Small colored
  arrows on the energy axis mark the highest occupied levels.
  The curves are aligned with respect to the valence band and the energy
  of the bulk VB edge is set to zero.
\label{fig:DOS:all}}
\end{figure}

The total density of states (DOS) of a representative Zn$_{32}$Mg$_4$O$_{36}$ 
sample calculated with the SIC-LDA approach is shown in Figure 
\ref{fig:DOS:all}. The upper part of the VB is mainly formed by the O 2p 
valence orbitals while the lower part is dominated by the Zn 3d semicore 
orbitals. Experimentally, the average d-band energy of ZnO was determined to be 
at -7.8 eV relative to the VB edge.\cite{he82} This is almost unchanged when Mg 
is introduced. The SIC-LDA approach reproduces this experimental observation in 
contrast to LDA, GGA and several commonly used hybrid functionals which place 
the Zn 3d levels too high in energy and therefore overestimate their 
hybridization with the O 2p levels. The latter can lead to wrong positions of 
defect levels in the band gap.\cite{ko14} Thus these results validate our approach to analyze individual electronic levels of point defects by using the 
SIC-LDA approach.

\subsection{Electronic defect levels near the VB}

\begin{figure}[]
  \centerline{\includegraphics[width=8.5cm]{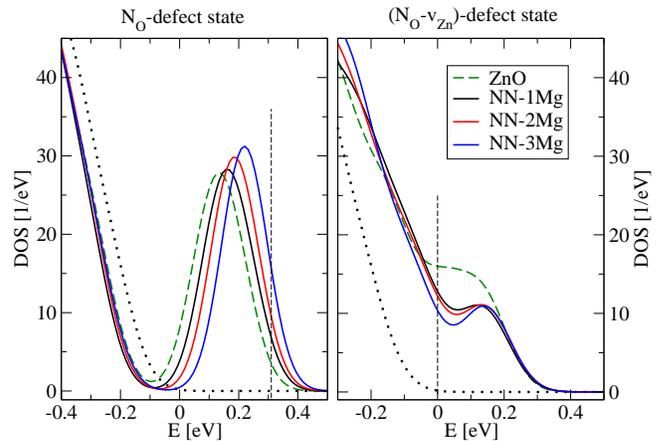}}
\caption{(Color online)
Section of the total DOS in the vicinity of the VB edge of 
Zn$_{32}$Mg$_4$O$_{36}$ doped with N$_{{\rm O}}$ (left) and N$_{{\rm O}}$ + 
v$_{{\rm Zn}}$ (right), calculated with the SIC-LDA, as function of the number 
of nearest-neighbor Mg atoms. For comparison, results for defects in ZnO are 
shown as dashed lines and the DOS of perfect ZnO is included as a dotted line.
\label{fig:defects_mg}}
\end{figure}

From the total density of states (DOS) in Fig.\ \ref{fig:DOS:all} it can be 
seen that the defects N$_{{\rm O}}$, v$_{{\rm Zn}}$ and N$_{{\rm O}}$-v$_{{\rm 
Zn}}$ create energy levels close the VB edge. However, a closer inspection -- 
which is provided in the inset -- shows that the defect levels that originate 
from N$_{{\rm O}}$ are separated from the VB with a peak at about 0.25 eV.
In the left panel of Fig.\ \ref{fig:defects_mg} the position of the N$_{{\rm
O}}$ defect level is shown as function of the number of nearest Mg neighbors
(NN). Here, NN-1Mg indicates a Zn$_{32}$Mg$_4$O$_{35}$N (or 
Zn$_{31}$Mg$_4$O$_{35}$N) supercell in which one Mg atom is adjacent to the 
N$_{\rm O}$ (or N$_{\rm O}$-v$_{\rm Zn}$) defect, respectively, while in the 
NN-2Mg and NN-3Mg supercells,  two and three Mg atoms are nearest neighbors of 
the defect.

An increasing number of NN Mg atoms leads to an up-shift of the defect level. 
Therefore, concerning p-type conductivity, Mg doping makes N$_{\rm O}$ defects 
less attractive since the level becomes more narrow and moves deeper into the 
band gap. On the other hand, the energy level position of the N$_{\rm
O}$-v$_{\rm Zn}$ defect complex remains almost unchanged for different numbers 
of NN Mg atoms.

The substitution of Zn by Mg in ZnO does not help to make the acceptor levels 
of individual N$_{{\rm O}}$ defects shallow, which would facilitate p-type 
conductivity at room temperature, but rather provokes the contrary.
Moreover, it has been reported\cite{gai08} that the presence of Mg atoms in the neighborhood of a 
N$_{{\rm O}}$ defect increases the defect formation energy considerably and therefore reduces the respective defect concentration in the material. Our calculations are in line with the findings of Ref. \onlinecite{gai08}, see discussion in the Appendix.  

Zinc vacancy defects alone create very shallow defect levels (see inset of 
Fig.\ \ref{fig:DOS:all}), which is a well known result from previous 
studies.\cite{ko11b} Experimentally, defect levels at either 130meV\cite{re13}, 
160 meV\cite{ku14}, or 190 meV\cite{mu12}  are reported, depending on the 
specific experiment. According to Fig.\ \ref{fig:defects_mg}, the levels of the 
N$_{{\rm O}}$-v$_{{\rm Zn}}$ defect complex lie higher above the VB edge than 
that of the individual v$_{{\rm Zn}}$ defect and fit to the experimentally 
determined values rather well.
We thus propose to assign the experimentally found acceptor levels to a
N$_{{\rm O}}$-v$_{{\rm Zn}}$ defect complex. This complex is very promising
since it provides shallow defect levels that are not separated in energy from the
VB edge, and thus continuous low energy excitation are possible. The position of 
the defect level varies only slightly with the number of Mg atoms near the 
defect complex (see right panel of Fig.\ \ref{fig:defects_mg}).

\begin{figure}[]
\includegraphics[width=8.5cm,draft=false]{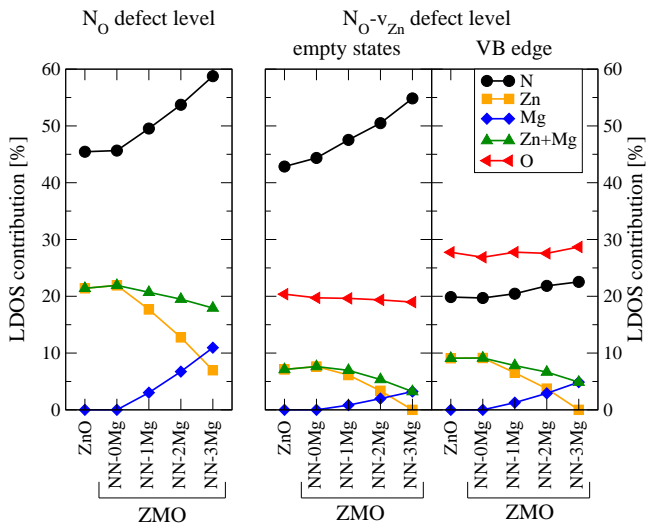}
\caption{(Color online)
Relative LDOS contributions to the total DOS of the N$_{\rm O}$-defect level 
(left), the unoccupied part of the (N$_{\rm O}$-v$_{\rm Zn}$)-defect level 
(middle) and the respective VB edge (right), shown for the N$_{\rm O}$ atom and 
its most relevant neighbors. The relative positions of these atoms, which are 
the ones most involved in the defect complex, are sketched on the left in Fig.\ 
\ref{fig:positions}. For the oxygen and metal ions the summed contributions are 
shown. \label{fig:defects} }
\end{figure}

\begin{figure}[]
\begin{center}
\includegraphics[width=7.5cm,draft=false]{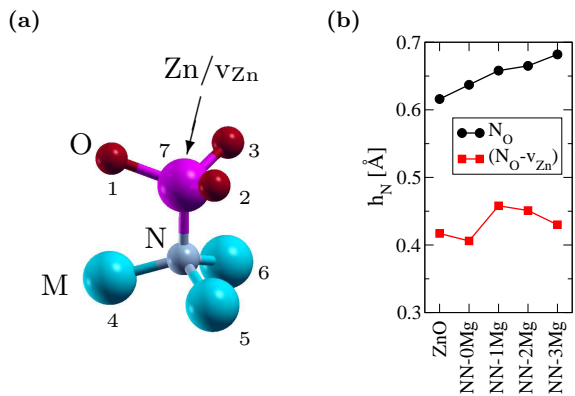}
\caption{(Color online)
(a) Sketch of the  most relevant atoms involved in the defects N$_{{\rm O}}$ 
and N$_{{\rm O}}$ + v$_{{\rm Zn}}$. Metal atoms on positions 4--6 can be either 
Zn or Mg. Position 7 corresponds to either a Zn atom or a Zn vacancy. (b) 
Dependency of the N atom position (height h$_N$) above the atomic plane defined 
by the metal atoms 4--6 on the number of Mg atoms involved.
\label{fig:positions}}
\end{center}
\end{figure}

\subsection{Spatial distribution of N$_{\rm O}$ and N$_{\rm
O}$-v$_{\rm Zn}$ defect levels}

In order to analyze the spatial distribution of the defect levels we have
evaluated the atom-projected local densities of states (LDOS) for the most 
relevant atoms involved in the defect (complex).
Figure\ \ref{fig:defects} displays their contributions to the defect levels 
while the relative locations of these atoms are sketched in Fig.\ 
\ref{fig:positions}a. A contribution of an individual neighbor atom to a 
defect level is evaluated by integrating the LDOS of the respective atom in the 
relevant energy interval and subsequent division by the sum of the LDOS of all 
atoms in the supercell, integrated for the same energy range. The energy 
interval [-0.1, 0.5]eV was chosen for the analysis of the N$_{\rm O}$ 
defect, while the analysis of the N$_{\rm O}$-v$_{\rm Zn}$ defect complex is 
carried out separately for the intervals [-0.2, 0.0]eV (VB edge) and  
[0.0, 0.4]eV (unoccupied states above the VB edge), c.f. Fig. 
\ref{fig:defects_mg}.

Note that the contributions for the oxygen ions (red line) and metal ions (green 
line) shown in Fig.\ \ref{fig:defects} are the sums for the 3 O-atoms (numbered 
1--3 in Fig.\ \ref{fig:positions}a) and the 3 or 4 metal atoms (numbered 4--7 
in Fig.\ \ref{fig:positions}a), respectively. The metal atoms at positions 4--6 
can be either Zn or Mg, while position 7 corresponds to either a Zn atom in the 
case of the N$_{\rm O}$ defect or to a Zn vacancy, in the case of the N$_{\rm 
O}$-v$_{\rm Zn}$ defect complex.

The N$_{\rm O}$ defect level is highly localized on the central N-atom, by
more than $45\%$, as can be seen from the left panel in Fig.\ \ref{fig:defects}.
About $20\%$ is shared by the four NN metal atoms, while the oxygen atoms of the 
second neighbor shell contribute with $2\%$ or less each.
The localization on the N-atom increases to $\sim60\%$ with an increasing 
number of NN Mg-atoms, while the total contribution of the NN metal atoms 
remains rather constant. This increase in localization is connected to an 
increasing spatial separation of the N atom from the plane defined by the three 
metal atoms 4--6 with increasing number of Mg atoms therein, see graph in Fig. 
\ref{fig:positions}b.

The analysis of the spatial distribution of the N$_{\rm O}$-v$_{\rm Zn}$ defect 
(c.f. middle and right panel  in Fig.\ \ref{fig:defects}) is divided into two 
parts, namely (i) the unoccupied states above the VB edge and (ii) the VB edge 
which is modified with respect to the bulk material by the presence of the
defect. The unoccupied states (middle panel in Fig.\ \ref{fig:defects}) are
reminiscent of the N$_{\rm O}$ defect and the N atom contributes more
than 40\% in pure ZnO which increases up to 55\% with increasing number of NN
Mg. The hybridization with the v$_{\rm Zn}$ defect state results in a $20\%$
contribution of the three oxygen atoms next to the vacancy site. The three
remaining NN metal atoms (numbered 4--6 in Fig.\ \ref{fig:positions}a) only 
contribute $\approx$2--3\% (Zn) and $\approx$1\% (Mg),
each. For the VB edge (right panel in Fig.\ \ref{fig:defects}), reminiscent of 
the v$_{\rm Zn}$ defect, the 3 oxygen atoms are more important ($\approx$27\%), 
but the N-atom still contributes on average with $\approx$20\%. A change in the 
number of NN Mg atoms has only little effect on this spatial distribution with 
a small shift of weight from the NN metal atoms to N.

\subsection{Defect-complex mechanism for p-type conduction mechanism}

For p-type conduction a Fermi level close the VB edge and shallow acceptor 
levels in sufficiently high concentration are needed.
Since the thermal excitation energy at room temperature is only about 25 meV,
p-type conduction with a significant hole concentration can at best be achieved
by acceptor defect levels which are distributed continuously right above the VB 
edge. Our SIC-LDA analysis of the most relevant defects in N-doped ZMO shows
that v$_{{\rm Zn}}$ and N$_{{\rm O}}$-v$_{{\rm Zn}}$ indeed provide such
defect levels.

Our results support the hypothesis that co-doping of ZnO with Mg and N can be a possible
route to obtain p-type conductivity. The (N$_{{\rm O}}$-v$_{{\rm Zn}}$) defect complex 
yields shallow acceptor levels which can be activated by the experimentally reported lowering of  
the net electron concentration in ZMO.\cite{Li07,Qiu07,hi09} This carrier compensation effect increases linearly with the Mg content and can be explained by deep-lying Mg-related defects which 
act as charge traps.\cite{hi09}

Zinc vacancies  are low lying energy defects which exist also in 
pure ZnO.\cite{ko11b} However, pure ZnO usually is an intrinsically n-type 
semiconductor since donor defects dominate. Only grain boundaries in 
polycrystalline ZnO may provoke p-type conducting films\cite{wa09} under oxygen 
rich conditions. In general the addition of oxygen is thought to reduce the 
number of oxygen vacancies which are supposed to be the key donor defects.
We believe that an understanding of the effect of oxygen addition has to be 
extended to take the grain boundaries into account.
Under oxygen rich conditions the percentage of grain boundaries with
undercoordinated oxygen atoms increases. In previous work\cite{ko10,ko11}
we have shown that such grain boundaries contribute shallow levels above the VB
edge. These are an equivalent to the defect levels of v$_{{\rm Zn}}$ in 
bulk-like ZnO or ZMO discussed here, since the oxygen atoms at the 
grain boundaries are undercoordinated due to local Zn deficiencies. 
Furthermore, nitrogen at grain boundaries can form shallow defect levels  which 
are very similar to those shown in  Fig.\ \ref{fig:defects_mg} (see Fig. 11 in 
Ref. \onlinecite{ko10}).
We argue therefore that all these shallow levels are activated due to
the depletion zones near grain boundaries.\cite{wa09,al06}
Zhang et al. interpret their experimental results along this line,\cite{zha06}
and Wang et al. even suppose that oxygen excess causes the
formation of a quasi two dimensional hole gas which could explain the very high
mobilities seen in some experiments (see Ref. \onlinecite{wa09} and references
therein).

Putting all from above together, this rather general defect-complex mechanism 
for promoting p-type conductivity in ZnO, ZMO, and related materials emerges: 
defect complexes which are connected to the zinc vacancies in the bulk interior 
or at grain boundaries provide shallow defect levels which can serve as 
acceptor defects. The activation of these defects by lowering the net electron 
concentration is provided either by co-doping (e.g. N and Mg\cite{hi09}) or by depletion 
zones around grain boundaries.

\section{Summary}

We have studied the electronic densities of states of crystalline ZnO and 
Zn$_{1-x}$Mg$_x$O with $0 \le x \le 0.4$. We have shown that the SIC-LDA allows
to calculate the electronic band gap as function of the Mg content
and leads to quantitatively reasonable results with respect to 
experimental data.

We have determined the electronic defect levels which are generated by the
introduction of N$_{{\rm O}}$ substitutional defects. They are
at 0.25 eV above the VB or higher. An increasing number of neighboring
Mg atoms to the N$_{{\rm O}}$ site shifts the defect level even deeper into the
band gap. On the other hand, isolated v$_{{\rm Zn}}$ defects generate shallow
levels attached to the VB edge. In combination with N$_{{\rm O}}$, the defect 
complex N$_{{\rm O}}$-v$_{{\rm Zn}}$ produces low lying defect
levels which emerge continuously above the VB and no detached gap states (like
in the case of the isolated N$_{{\rm O}}$ defect) appear. It is likely
that these levels which are peaked at 125--140 meV are connected to 
experimentally determined acceptor levels which are reported at 130 
meV\cite{re13}, 160 meV\cite{ku14} or 190 meV.\cite{mu12} Neighboring Mg atoms 
to the N$_{{\rm O}}$-v$_{{\rm Zn}}$ defect complex do not change the situation 
significantly. Apparently the role of Mg is mainly a lowering of the net 
electron concentration which lowers the Fermi level and therefore contributes 
the N$_{{\rm O}}$-v$_{{\rm Zn}}$ defect complexes as acceptors.

Our presented results are consistent with those of an earlier study on ZnO 
grain boundaries\cite{ko10,ko11}, in which we found that at
grain boundaries with local Zn deficiencies (which is equivalent to
undercoordinated oxygen atoms), N$_{{\rm O}}$ defects at
the grain boundary produce shallow defect levels without gap to the VB edge
and are thus promising acceptor levels.

The following conclusion emerges from this work:
one condition for p-type conductivity in ZnO, ZMO, or related materials is to
have enough p-type defects like for example N$_{{\rm O}}$ which interact with 
v$_{{\rm Zn}}$ point defects, or local Zn deficiencies in extended defects like 
grain boundaries, in order to form shallow gapless defect levels above the VB 
edge. Moreover, the net electron concentration has to be low enough
in order to make these shallow acceptor states partially unoccupied.
This second condition can be achieved by Mg$_{{\rm Zn}}$/N$_{{\rm O}}$ 
co-doping in the bulk interior or by N$_{{\rm O}}$ doping at grain boundaries 
to create depletion zones.

\section{Acknowledgments}

Financial support for this work was provided by the European
Commission through contract No. NMP3-LA-2010-246334 (ORAMA).

\section{Appendix}

\begin{figure}[]
\begin{center}
\includegraphics[width=\columnwidth,draft=false]{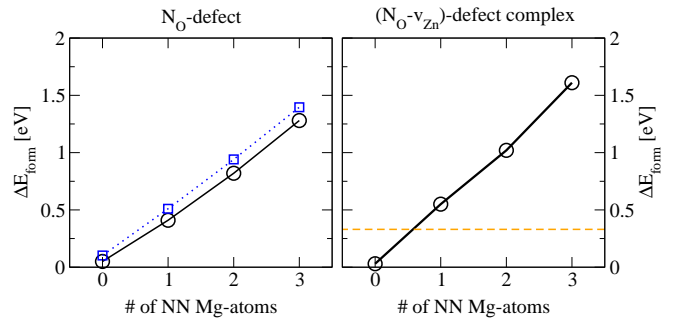}
\caption{(Color online)
\label{fig:Eform}
Increase of the defect formation energy of the N$_{{\rm O}}$ defect (left) and the N$_{{\rm O}}$-v$_{{\rm Zn}}$ complex (right) with increasing number of next neighbor Mg atoms. The energy differences $\Delta E_{{\rm form}}$ are calculated with respect to the values obtained for bulk ZnO. Results from Ref.  \onlinecite{gai08} are shown as a dotted line for comparison in the left panel. The sum of the formation energies of isolated N$_{{\rm O}}$ and v$_{{\rm Zn}}$ (without next neighbor Mg atoms) is shown as a dashed line in the right panel.}
\end{center}
\end{figure}

\subsection{Defect formation energies}
The defect formation energies of the zinc vacancy and the N$_{{\rm O}}$ substitutional defect in ZnO  have recently been discussed in great detail (see e.g. Refs. [\onlinecite{Jan07,Oba08,Cla10,
ko10,ko11c,am15,sak13}] and Refs. therein). For a discussion of defect formation energies in ZnO obtained from SIC-LDA we refer the reader to Refs. [\onlinecite{ko11,ko11c}].
Here we want to focus on the effect of the Mg atoms on the defect formation energies (E$_{{\rm form}}$) of N$_{{\rm O}}$ and the N$_{{\rm O}}$-v$_{{\rm Zn}}$ complex. E$_{{\rm form}}$ is found to depend strongly on the number of Mg atoms being nearest neighbor to the defect site, which can be explained by a stronger Zn-N bond compared to the Mg-N bond. Figure \ref{fig:Eform} shows the increase in E$_{{\rm form}}$ with respect to bulk ZnO with an increasing number of next neighbor Mg atoms. Our results are in line with the findings of Gai \emph{et al}\cite{gai08} who have performed DFT simulations using cubic zinc-blende ZMO alloys. 
Note that the formation of a N$_{{\rm O}}$-v$_{{\rm Zn}}$ complex is energetically favorable by $~0.3eV$ compared to a pair of isolated N$_{{\rm O}}$ and v$_{{\rm Zn}}$ point defects without Mg atoms in direct vicinity of the N atom.

\subsection{Supercell size effects}

\begin{figure}[]
\begin{center}
\includegraphics[width=7cm,draft=false]{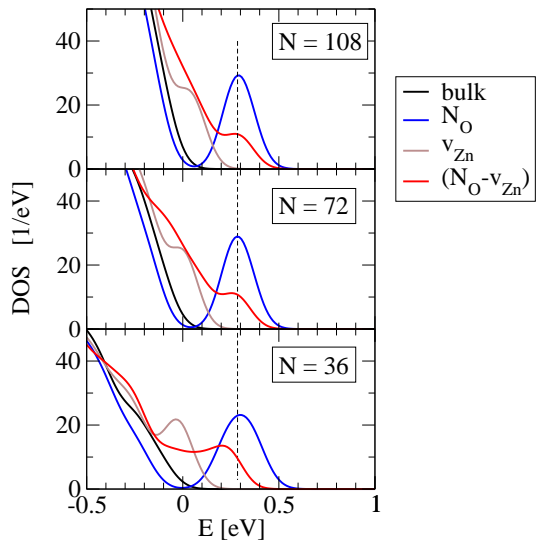}
\caption{(Color online)
\label{fig:supercellsize}
Zoom on the DOS in vicinity of the valence band edge, showing the defect levels of N$_{{\rm O}}$, v$_{{\rm Zn}}$, and the (N$_{{\rm O}}$-v$_{{\rm Zn}}$)-complex calculated for three different supercell sizes.
}
\end{center}
\end{figure}

In order to check the influence of the finite supercell size on the defect levels obtained from our DOS calculations, we have compared calculations for three different supercell sizes, namely N=36, 72 and 108 atoms. As shown in Fig. \ref{fig:supercellsize}, there is only very little change in level position and peak shape, when comparing the results for 72 and 108 atom supercells. On the other hand, the 36-atom supercells do not yet yield fully converged results, although the general conclusions could qualitatively also be drawn from these rather small cells. We therefore decided to use the 72-atom supercells for the results presented in the main part of the paper.


\begin{thebibliography}{99}


\bibitem[]{le01} E.-C. Lee, Y.-S. Kim, Y.-G. Jin and K. J. Chang, 
Phys. Rev. B \textbf{64}, 085120 (2001).

\bibitem[]{le04} E.-C. Lee and K. J. Chang, 
Phys. Rev. B \textbf{70}, 115210 (2004).

\bibitem[]{le06} W.-J. Lee, J. Kang and K. J. Chang, 
Phys. Rev. B \textbf{73}, 024117 (2006).

\bibitem[]{pa02} C. H. Park, S. B. Zhang and S. H. Wei, 
Phys. Rev. B \textbf{66}, 073202 (2002).

\bibitem[]{li04} S. Limpijumnong, S. B. Zhang, S.-H. Wei and C. H. Park, 
Phys. Rev. Lett. \textbf{92}, 155504 (2004).

\bibitem[]{ya08}  Y. Yan and S.-H. Wei,
Phys. Stat. Sol. \textbf{245}, 641  (2008).

\bibitem[]{wa03} L. G. Wang and A. Zunger, 
Phys. Rev. Lett. \textbf{90}, 256401 (2003).

\bibitem[]{ya07} Y. Yan, J. Li, S. H. Wei and M. M. Al-Jassim, 
Phys. Rev. Lett. \textbf{98}, 135506 (2007).

\bibitem[]{zha06} C. Y. Zhang, X. M. Li, X. D. Gao, J.L. Zhao, K. S.  and J. M. Bian, 
Chem. Phys.  Lett. \textbf{420}, 448 (2006).
%codoping N/Al

\bibitem[]{ly09}   J. L. Lyons, A. Janotti and C. G. Van de Walle,
Appl. Phys. Lett. \textbf{95}, 252105 (2009).

\bibitem[]{la10} S. Lany and A. Zunger, 
Phys. Rev. B \textbf{81}, 205209 (2010).

\bibitem[]{ko10} W. K\"orner and C. Els\"asser, 
Phys. Rev. B \textbf{81}, 085324 (2010).

\bibitem[]{re13} J. G. Reynolds, C. L. Reynolds Jr., A. Mohanta, J. F. Muth,
J. E. Rowe, H. O. Everitt and D. E. Aspnes, 
Appl. Phys. Lett. \textbf{102}, 152114 (2013).

\bibitem[]{liu12} L. Liu, J. Xu, D. Wang, M. Jiang, S. Wang, B. Li,
Z. Zhang, D. Zhao, C.-X. Shan, B. Yao and D. Z. Shen, 
Phys.  Rev. Lett. \textbf{108}, 215501 (2012).

\bibitem[]{am15} M. N. Amini,  R. Saniz, D. Lamoen  and B. Partoens, 
Phys. Chem. Chem. Phys. \textbf{17}, 5485 (2015).

\bibitem[]{wa09} B. Wang, J. Min, Y. Zhao,W. Sang and C. Wang,
Appl. Phys. Lett. \textbf{94}, 192101 (2009).
% depletion picture , oxygen dopin at gb

\bibitem[]{hi09} A. Hierro, G. Tabares, J. M. Ulloa, E. Mu\~noz,
A. Nakamura, T. Hayashi and J. Temmyo,
Appl. Phys. Lett. \textbf{94}, 232101 (2009).
% carrier compensation, mg content

\bibitem[]{ko11} W. K\"orner and C. Els\"asser, 
Phys. Rev. B \textbf{83}, 205315 (2011).

\bibitem[]{ko14} W. K\"orner, D. F. Urban, D. M. Ramo,
P. D. Bristowe and C. Els\"asser,
Phys. Rev. B \textbf{90}, 195142 (2014).

\bibitem[]{sa14} S. Sallis, K. T. Butler, N. F. Quackenbush, D. S. Williams, M. Junda, D. A. Fischer, J. C. Woicik, N. J. Podraza, B. E. White, Jr.,  A. Walsh and L. F. J. Piper, 
Appl. Phys. Lett. \textbf{104}, 232108 (2014).

\bibitem[]{sa15} S. Sallis, N. F. Quackenbush, D. S. Williams, M. Senger,
J. C. Woicik, B. E. White and L. F. J. Piper, 
Phys. Stat. Sol. (a) \textbf{212}, 1471 (2015).

\bibitem[]{ko13} W. K\"orner, D. F. Urban and C. Els\"asser, 
J. Appl. Phys. \textbf{114}, 163704 (2013).

\bibitem[]{ko15} W. K\"orner, D. F. Urban and C. Els\"asser, 
Phys. Stat. Sol. (a) \textbf{212}, 1476 (2015).

\bibitem[]{ku14} A. Kurtz, A. Hierro, E. Mu\~noz, S. K. Mohanta,
A. Nakamura, and J. Temmyo, 
Appl. Phys. Lett. \textbf{104}, 081105 (2014).
% znmgo:N levels

%%%%%%%%%%%%%%%%%%%%%%%%%%%%%%%%%%%%%%%%%%%%%%%%%%%%%%%%
% theorie papers computational details
%\bibitem[]{pe96} J. P. Perdew,  K. Burke  and M. Ernzerhof, Phys. Rev. Lett.
%\textbf{77}, 3865 (1996).


\bibitem[]{bl94} P. E. Bl\"ochl, 
Phys. Rev. B \textbf{50}, 17953 (1994).

\bibitem[]{kr96} G. Kresse and J. Furthm\"uller, 
Phys. Rev. B \textbf{54}, 11169 (1996).

\bibitem[]{kr99} G. Kresse and D. Joubert, 
Phys. Rev. B \textbf{59}, 1758 (1999).

\bibitem[]{ro81} J. H. Rose, J. Ferrante and J. R. Smith, 
Phys. Rev. Lett. \textbf{47}, 675 (1981).

\bibitem[]{el90} C. Els\"asser, N. Takeuchi, K. M. Ho, C. T. Chan
P. Braun and M. F\"ahnle, 
\mbox{J. Phys.:} Condens. Matter \textbf{2}, 4371 (1990).

\bibitem[]{el92} K. M. Ho, C. Els\"asser, C. T. Chan and M. F\"ahnle, 
\mbox{J. Phys.:} Condens. Matter \textbf{4}, 5189 (1992).

\bibitem[]{mey95} B. Meyer, K: Hummler, C. Els\"asser and M. F\"ahnle,
\mbox{J. Phys.:} Condens. Matter \textbf{7}, 9201 (1995).

\bibitem[]{lech04} F. Lechermann, M. F\"ahnle, B. Meyer and  C. Els\"asser,
Phys. Rev. B \textbf{69}, 165116 (2004).

\bibitem[]{ko11b} W. K\"orner, P. D. Bristowe and C. Els\"asser,
Phys. Rev. B \textbf{84}, 045305 (2011).
% ZnO gbs

\bibitem[]{ko11c} W. K\"orner and C. Els\"asser, 
Phys. Rev. B \textbf{83}, 205306 (2011).
% si-nb zno

\bibitem[]{pe81} J. P. Perdew and  A. Zunger, 
Phys. Rev. B \textbf{23}, 5048 (1981).

\bibitem[]{va85} D. Vanderbilt, 
Phys. Rev. B \textbf{32}, 8412 (1985).

\bibitem[]{vo96} D. Vogel, P. Kr\"uger and J. Pollmann,
Phys. Rev. B \textbf{54}, 5495 (1996).

%%%%%%%%%%%%%%%%%%%%%%%%%%%%%%%%%%%%%%%%%%%%%%%%%%%%%%5

%\bibitem[]{de03} F. Decremps, F. Datchi, A. M. Saitta, A. Polian,
%S. Pascarelli, A. Di Cicco, J. P. Itie and F. Baudelet
% Phys. Rev. \textbf{B} 68, 104101 (2003).

\bibitem[]{ry06} H. Ryoken, N. Ohashi, I. Sakaguchi, Y. Adachi,
S. Hishita and H. Haneda,
J. Cryst. Growth \textbf{287}, 134 (2006).
% Bandgaps values

\bibitem[]{gh07} R. Ghosh and D. Basak,
J. Appl. Phys. \textbf{101}, 023507 (2007).
% Bandgaps values

\bibitem[]{kau09} A. Kaushal and D. Kaur
Solar Energy Materials \& Solar Cells \textbf{93}, 193 (2009).
% bandgap values and lattice parameters, experiment

\bibitem[]{nish10} H. Nishinaka1, Y. Kamada, N. Kameyama, and S. Fujita,
Phys. Stat. Sol. (b) \textbf{247}, 1460 (2010).
% lattice parameters, experiment

\bibitem[]{lau13} B. Laumer, F. Schuster, M. Stutzmann, A. Bergmaier,
G. Dollinger, and M. Eickhoff
J. Appl. Phys. \textbf{113}, 233512 (2013).
% lattice parameters, experiment

\bibitem[]{franz13} C. Franz, M. Giar, M.Heinemann, M. Czerner, and C. Heiliger,
MRS Proceedings  \textbf{1494}, 57 (2013).
% lattice parameters, DFT

\bibitem[]{gu12} E. G\"ur, G. Tabares, A. Arehart, J. M. Chauveau, A. Hierro,
and S. A. Ringel,
J. Appl. Phys. \textbf{112}, 123709 (2012).
% Bandgaps values

\bibitem[]{te00} C. W. Teng, J. F. Muth, U. Ozgur, M. J. Bergmann, H. O.
Everitt, A. K. Sharma, C. Jin, and J. Narayan,
Appl. Phys. Lett. \textbf{76}, 979 (2000).
% Bandgaps values

\bibitem[]{pa01} W. I. Park, G.-C. Yi, and H. M. Yang,
Appl. Phys. Lett. \textbf{79}, 2022 (2001).
% Bandgaps values


\bibitem[]{he82} \emph{Semiconductors: Physics of Group 4 Elements
and 3-5 Compounds}, edited by K. H. Hellwege and O. Madelung,
Landbolt-B\"ornstein, New Series, Group 3, Vol. 17, Pt. A (Springer,
Berlin 1982); \emph{Semiconductors: Intrinsic Properties of Group 4
Elements and 3-5-7, and 1-7 Compounds}, edited by 
O. Madelung, W. von der Osten, and U. R\"ossler,
Landolt-B\"ornstein, New Series, Group 3, Vol 22, Pt. A
(Springer, Berlin 1987).
% d-band ZnO

\bibitem[]{gai08} Y. Q. Gai, B. Yao, Z. P. Wei, Y. F. Li, Y. M. Lu, D. Z. Shen, J. Y. Zhang,
D. X. Zhao, X. W. Fan, Jingbo Li, and Jian-Bai Xia,
Appl. Phys. Lett. \textbf{92}, 062110 (2008).

\bibitem[]{mu12} P. Muret, D. Tainoff, C. Morhain and J. M Cauveau,
 Appl. Phys. Lett. \textbf{101}, 122104 (2012).
% znmgo:N levels

\bibitem[]{Li07} Y. F. Li, B. Yao, Y. M. Lu, Z. P. Wei, Y. Q. Gai, C. J. Zheng, Z. Z. Zhang, 
B. H. Li, D. Z. Shen, X. W. Fan, and Z. K. Tang,
Appl. Phys. Lett. \textbf{91}, 232115 (2007).

\bibitem[]{Qiu07} M. X. Qiu, Z. Z. Ye, H. P. He, Y. Z. Zhang, X. Q. Gu, L. P. Zhu, and B. H. Zhao,
Appl. Phys. Lett. \textbf{90}, 182116 (2007).

\bibitem[]{al06} M. A.  Alim, S. Li, F. Liu and P. Cheng, 
Phys. Stat. Sol. (a) \textbf{203}, 410 (2006).
%depletion zone grain boundaries

\bibitem[]{Cla10} S. J. Clark, J. Robertson, S. Lany, and A. Zunger,
Phys. Rev. B \textbf{81}, 115311 (2010).

\bibitem[]{Jan07} A. Janotti and C. G. Van de Walle,
Phys. Rev. B \textbf{76}, 165202 (2007).

\bibitem[]{Oba08} F. Oba, A. Togo, I. Tanaka, J. Paier, and G. Kresse,
Phys. Rev. B \textbf{77}, 245202 (2008).

\bibitem[]{sak13} S. Sakong, J. Gutjahr an P. Kratzer, 
J. Chem. Phys. \textbf{138}, 234702 (2013).

\end{thebibliography}
\end{document}